# Engineering Quantum Confinement in Semiconducting van der Waals Heterostructure

K. Wang[1], T. Taniguchi[2], K. Watanabe[2], P. Kim[1] *

[1]Department of Physics, Harvard University, Cambridge, 02138, MA, US

[2]National Institute for Materials Science, Namiki 1-1, Ibaraki 305-0044, Japan

*Email: pkim@physics.harvard.edu

**Spatial confinement and manipulation of charged carriers in semiconducting nanostructures are essential for realizing quantum electronic devices [1-3]. Gate-defined nanostructures made of two-dimensional (2D) semiconducting transition metal dichalcogenides (TMDCs) have the potential to add a unique additional control of quantum degrees of freedom owing to valley-spin locking of confined carriers near the band edges [4-13]. However, due to prevailing inhomogeneities in the conducting channels, it has been difficult to realize quantum confinement in 2D TMDCs with well-controlled tunnel-coupling strength [14-16]. Here we demonstrate quantum transport in lateral gate-defined 2D electron quantum dots formed in atomically thin TMDC heterostructures. Utilizing micro-fabricated local contact gates, encapsulation in 2D dielectrics and light illumination at low temperatures, we show that the quality of TMDC 2D electron gases (2DEGs) can be improved, rendering them suitable for mesoscopic quantum transport measurements. We observe quantized conductance in quantum point contact (QPC) channels controlled by gate-tunable confinement. We also demonstrate single electron transport in TMDC quantum dots (QD) with tunable tunnel-coupling. Our observation holds promise for the quantum manipulation of spin and valley degrees of freedom in engineered TMDC nanostructures, enabling versatile 2D quantum electronic devices.**

Recently, TMDC semiconductors have attracted much research interest due to their unique band structures and material properties [4-16]. The in-plane hexagonal lattice structure with two inequivalent lattice sites generate band edges found at the corner of the Brillouin zone, known as two inequivalent K and K' valleys. The broken inversion symmetry of the lattice in combination with the strong spin-orbit interaction polarizes spins in the opposite direction near the band edges of each valley. This combined spin-valley degree of freedom has been proposed as the key ingredient in a number of versatile novel quantum electronic and optoelectronic devices, such as valley-spin qubits [10], localized excitons and trions [17], and electrostatically coupled low-dimensional drag devices [18-19].

The first step toward quantum manipulation of electrons is to demonstrate charge confinement with tunable tunnel-couplings [1-3]. Previously, quantum dots have been demonstrated in the 2D limit TMDCs in top-gated $WSe_2$ and $WS_2$ on $SiO_2$ substrate [15,16]. However, the carrier mobility and 2DEG homogeneity is significantly limited due presumably to the charge traps originated from the $SiO_2$ substrate and the surface adsorption remained from the device fabrication process. The resulting inhomogeneous 2DEG can lead to accidental quantum dot formation and thus significantly reduced device controllability. Similar problems were identified in mesoscopic quantum devices including modulation doped Si/SiGe QDs [20], short $MoS2$ channels [14] and graphene nanoribbons [19,21]. Although single electron tunneling (SET) behavior can be demonstrated in these accidental QDs, tunnel couplings are difficult to control due to their disorder-defined nature. Therefore, improving the quality and homogeneity of the 2DEG is necessary in order to fabricate tunable TMDC quantum devices.

To achieve high mobility TMDC channel, we employ van der Waals (vdW) heterostructure consist of $hBN/MoS_2/hBN$ with graphene vdW contacts [12], assembled using a dry transfer technique [22] (Fig. 1a). Prior to assembly, graphene samples are chemically n-doped to provide enough molecular adsorbates to facilitate work function alignment between graphene and $MoS_2$. The resulting vdW heterostructure consists of two pieces of graphene providing vdW contacts to a piece of intrinsic MoS$_2$ channel (trilayer, ~2nm thick), which are then encapsulated in hBN (60-80 nm). The heterostructure is then transferred on top of pre-deposited Cr/PdAu (1nm/5nm) local gate structures and subsequently annealed at 350 C for 15 minutes. The back gate structure contains a pair of large local gates near the contacts, covering the region where graphene and MoS$_2$ overlap. These gates help achieve good contact via field-induced doping without compromising the low carrier density in the QD region. The fine gates in the center region (Fig. 1b inset) are used to define the quantum electrostatic confinement potential. The entire device processing, including exfoliation, atomic force microscopy sample characterization, transferring and annealing, are done in a glove box with Ar gas-filled environment to minimize contamination and deterioration during the sample preparation process. After the assembly is complete, the vdW heterostructure is taken out of the glove box and 1D edge contacts [22] to each graphene layers are fabricated with Cr/Pd/Au (1.5nm/5nm/120nm) metal contacts.

We characterize the mobility and 2DEG homogeneity with Hall bar devices fabricated on the heterostructures described above. A voltage applied to the silicon back gate ($V_{BG}$) is used to tune the bulk carrier density while local gates near the contacts are biased at 10V to minimize the contact barriers. We also employ short period of (for ~20s) light illumination using an infrared (IR) - light emitting diode (LED) at low temperatures, which has been previously associated with improved 2DEG mobility [20, 23]. Figure 2a shows the zero-field longitudinal resistance ($R_{xx}$) as a function of $V_{BG}$ before and after LED illumination, showing the n-type field effect transistor (FET) behavior. A linear carrier density tunability ($n$=3-7x10$^{12}$ cm$^{-2}$,) with electron mobility $\mu$~6,000 cm$^2$/Vs, is demonstrated by a measurement performed at 1.7 K (Fig. 2b). Interestingly, we find that the threshold gate voltage shifts toward more positive voltage after LED illumination (from 17V to 22V), while the mobility and carrier density stays relatively unchanged (Fig. 2b). The underlying mechanism of the effect of LED illumination at low temperatures can be complicated at the microscopic level, and therefore difficult to be fully explored with transport measurements alone. A possible qualitative picture could be that the low energy IR photons, while unable to excite charge across the full band gap, are capable of facilitating local charge redistribution that potentially leads to a more homogeneous 2DEG (see SI).

The high mobility and homogeneity realized in our 2DEG can further be characterized by the density and magnetic field tunable Shubnikov de Haas (SdH) quantum oscillations in $R_{xx}$. As shown in Fig. 2c, a comprehensive SdH fan diagram can be constructed in the wide range of magnetic field and carrier density in the turn-on regime. A cross-section of the fan diagram at the low magnetic field regime exhibits regular SdH oscillation starting around 1T (Fig. 2d), implying a carrier mobility on the order of 10,000 cm$^2$/Vs, consistent with the value extracted from Hall measurements. The SdH peak near B=3T develops into two peaks (Fig. 2d), presumably corresponding to partial symmetry breaking. At high field (Fig. 2e), oscillations remain periodic in 1/B with higher frequency and more complicated features, which might originate in full symmetry-breaking of Landau levels and complex valley-spin dynamics near the band edges, as well as possible formation of parallel 2DEGs formed in each individual atomic layers [5,8]. While the exact nature of the residual-impurities-limited mobility in our sample is under investigation, a part of them originated from the contamination induced by the graphene contacts used in this study. With a new fabrication scheme similar to local doping by 'ion-implantation', the highest SdH mobility obtained in some of our devices using intrinsic MoS$_2$ channels reached 27,000 cm$^2$/Vs (see SI).

Employing these high-quality 2DEGs, we now can fabricate gate-defined mesoscopic structures with versatile gate control. We first discuss the configuration where all local gates are tied together to the same potential $V_{Local}$. (Fig. 3a) In this configuration, the current flows through gate-defined constrictions known as quantum point contacts (QPCs) [24]. In TMDC QPCs, quantized conductance in steps of 4e$^2$/h is expected considering the four-fold spin and valley degeneracy. Fig. 3a shows measured current across the device as a function of $V_{BG}$ and $V_{Local}$ with a 1 mV AC excitation and zero DC bias. As the carrier density becomes higher at more positive $V_{BG}$, the pinch-off happens at more negative $V_{Local}$, (with high on-off ratios of more than

$10^6$). 1D sweeps of $V_{Local}$ at several fixed back gate voltages exhibit plateau-like features (Fig. 3b), implying conductance quantization in QPC. Due to the comparably lower channel mobility and contact quality of the MoS$_2$ 2DEG than those of Si and GaAs counterparts, the two-probe conductance we measured across the QPC contains a significant contribution from the contact resistances and the bulk 2DEG series channel resistance leading to the QPC. Assuming the first plateau of conductance appears at $4e^2/h$, we can estimate this series resistance contribution $R_0$. The inset of Fig. 3b shows $R_0$ at different values of $V_{BG}$, where it is typically tens of kΩs and decreases with increasing $V_{BG}$, consistent with our measurements on contact resistances. We also note that at higher overall electron density (i.e. higher $V_{BG}$), the last plateau tends to split into two gradually with increasing 2DEG density implying the valley-spin locked band edges might be energetically resolved due to enhanced screening from charge inhomogeneities. Further improvement of the 2DEG and contact quality are required to demonstrate exact quantization, which may provide better insight into the valley-spin-locked band edges. We also note that the two series connected QPCs in our QD are not an ideal device configuration to demonstrate multiple conductance plateaus. Unlike multiple quantized conductance steps observed in our control device with symmetric split gates (Fig. S1a), the number of conductance channels can only be well-defined at low carrier densities in our QD geometry, and thus we observe conductance steps only near pinch-off of the QD coupling.

It is also important to note that the conductance of the device shown in Fig.3 exhibits a monotonic increase of G as a function of both $V_{BG}$ and $V_{QPC}$. This observation is in sharp contrast to 'accidental' quantum dot formation due to the disorder landscape, where rapid Coulomb oscillations are observed over a wide range of gate parameters. Such global presences of Coulomb oscillations in accidental quantum dots are due to the intrinsically fixed disorder-defined tunnel couplings, which are insensitive to applied gate voltages (Fig. S2b). The monotonic control of each QPC channel achieved in our gate defined QDs enables us to tune the QD-reservoir tunnel couplings precisely. Fig. 4a shows the measured current across the device as a function of $V_L$ and $V_R$ at fixed $V_N$=-8V and $V_C$=-10V. The conducting region is in a rectangular shape located at the upper right corner of the $V_R$-$V_L$ diagram, with plateaus whose locations depends only on a single corresponding gate voltage, demonstrating that the left (right) tunnel coupling is independently tuned by $V_L$ ($V_R$) as we designed. Near the lower-left corner of the conducting region (white dashed box), we found a suitable gate range where both tunnel couplings are optimized such that the tunnel-broadening of quantized levels in the dot becomes smaller than the charging energy $E_C$, while the current across device remains above the measurement threshold (~1pA).

With independent tunnel-coupling control of the QD, we can now demonstrate single-electron transistor (SET) operation of the device. Fig. 4b shows a detailed gate sweep in the region where we observe multiple regular oscillations of current across the device. Here, when one of the quantized levels of the quantum dot is in resonance with the Fermi energy, a finite tunnel current is measured. When no level is in resonance, the device is in the Coulomb blockade regime and current vanishes. The resonant condition is satisfied along negative slope lines in the $V_L$ vs $V_R$ parameter space, the slopes of which correspond to the ratio of their capacitive coupling to the QD, and whose separation is proportional to the charging energy $E_C$. The observation that the slope and the separation stay relatively constant over many consecutive charge transitions implies that the QD is uniformly defined. The slope of these transition lines is on the order of one, implying the resulting QD is located near the center of the device. We emphasize that these demonstrations are important metrics to differentiate the origin of the QD formation. If the QDs were accidentally formed due to 2DEG inhomogeneities, the narrowest constriction in the device, i.e., the QPC channels, would be a likely place. As a result, the resonant tunneling lines would appear horizontal or vertical in a $V_L$ vs $V_R$ sweep, since the capacitive couplings to the QDs are dominated by the gates that are defining the QPC channels (Fig S1c). In contrast, in a quantum dot defined in a uniform 2DEG, Coulomb blockade is observed in a much narrower range of gate parameters. Here, the device can be sensitively tuned from the tunnel-broadened regime (Δ-symbol in Fig. 4b) to the thermally-broadened regime (hexagon symbol in Fig. 4b) over small changes in gate voltages.

The charging energy of the QD can be measured by performing transport spectroscopy at a finite bias voltage $V_{SD}$. Figure 4c shows the current as a function of $V_{SD}$ while shifting the QD energy levels with a local gate voltage ($\Delta V_G$). A series of diamond-shaped zero-current regions in this diagram indicates the Coulomb blockade regime where no QD quantum level exists in the bias window defined by $V_{SD}$. The charging energy can be estimated from the height of the diamond along the $V_{SD}$ axis, where we obtain *Ec* ~ 2 meV. This value is in reasonable agreement with a dot size of r ~ 70nm, estimated according to a self-capacitance model [3]. This value is also consistent with the dot size expected from our gate design.

In conclusion, we have developed experimental methods to prepare homogeneous 2DEGs in $MoS_2$ van der Waals heterostructures, where periodic quantum oscillations are observed. With the high-quality 2DEG obtained, we have demonstrated essential signatures of quantum transport through both gate-defined 1-dimensional and 0-dimensional confinement nanostructures with sensitive tunability. The controlled charge localization demonstrated in this work provides a pathway toward control of the combined spin-valley degree of freedom in gate-defined structures on 2D TMDCs.

**Acknowledgements:**

We thank Jonah Waissman and Eunice Lee for helpful discussions. The major experimental work is supported by AFOSR (grant FA9550-14-1-0268). K. W. acknowledges support from ARO MURI (W911NF-14-1-0247). P.K. acknowledges partial support from the FAME Center, sponsored by SRC MARCO and DARPA. K.W. and T.T. acknowledge support from the Elemental Strategy Initiative conducted by the MEXT, Japan and JSPS KAKENHI Grant Numbers JP26248061, JP15K21722 and JP25106006. A portion of this work was performed at the National High Magnetic Field Laboratory, which is supported by National Science Foundation Cooperative Agreement No. DMR- 1157490 and the State of Florida. Nanofabrication was performed at the Center for Nanoscale Systems at Harvard, supported in part by an NSF NNIN award ECS-00335765.

**Author Contributions** KW performed the experiments and analyzed the data. KW and PK conceived the experiment. KW and PK wrote the paper. KW and TT provided hBN crystals.

**Author Information** Reprints and permissions information is available at xxxx. The authors declare no competing financial interests. Correspondence and requests for materials should be addressed to P.K. (e-mail: pkim@physics.harvard.edu).

**Figure legends**

**Figure 1. hBN-graphene-MoS$_2$ heterostructure with local backgates**. (a) The van der Waals heterostructure consists of two hBN (~60nm-80nm), two single layer graphene and one trilayer MoS$_2$. The stack has been assembled and transferred on top of the pre-patterned gates, using the dry-transfer technique. Subsequently, Cr/Pd/Au edge contacts are made to graphene flakes, making van der Waals electrical contact to the MoS$_2$ channel. (b) Optical microscope image of the gate-defined quantum dot

device. The inset shows a scanning electron microscope image. The scale bar is 200 nm. Graphene contacts to MoS$_2$ are improved by positively-biased local contact gates, while the quantum confinement in the quantum dot is defined by negatively-biased local fine gates.

**Figure 2. Homogeneous TMDC 2DEG with low carrier density.** (a) Metal-insulator transition of a Hall bar made on trilayer MoS$_2$ encapsulated in Ar environment. The turn-on threshold voltage shifts to more positive values after flashing the device with red LED at 1.7 K. (inset) Sample resistance at higher backgate voltages show no noticeable difference before and after LED flash. (b) A large range of gate-tunable carrier density and a Hall mobility up to ~ 6,000 cm$^2$/Vs has been achieved, independent of the use of LED flash. (c) With transparent contacts achieved at a relatively low bulk carrier density, measurement of Landau level fan diagram is possible. (d) Low-field SdH oscillation of device made with vdW ion-implantation method (see SI). The traces are displaced vertically for clarity of presentation. The oscillation starting at ~1T implies the SdH mobility of ~ 10,000 cm$^2$/Vs, consistent with Hall measurement. The SdH peak near B=3T develops into two peaks, corresponding to partial symmetry breaking. (e) High magnetic field data shows data with periodicity that is four times faster in 1/B, possibly implying full symmetry-breaking of Landau levels or parallel conducting channels from each atomic layers.

**Figure 3. Quantized conductance in quantum point contact.** (a) By energizing the quantum dot gates, two quantum point contact channels can be pinched off over a large range of electron density with a high on-off ratio exceeding 10$^6$. The conductance changes monotonically, implying a uniform QPC channel with straight-forward gate-control. (b) 1D cuts of the pinch-off curve at various $V_{BG}$ (from 30 V to 52.5 V, with an equidistance step of 1.5 V) reveal development of quantum plateau, a signature of quantum confinement. The inset shows the series resistance to the QPC estimated from the quantized plateaus.

**Figure 4. Gate-defined quantum dot in MoS$_2$**. (a) Zero-bias ac conductance across the device as a function of left gate voltage $V_L$ and right gate voltage $V_R$ at $V_N$ = -8V and $V_C$ = -10V. The rectangular shape of the conductance region indicates that the left (right) tunnel coupling getting tuned by $V_L$ ($V_R$) independently,

with visible quantum plateaus (inset). At the corner of the conducting region (dashed box), the tunnel couplings to the both contacts are optimized, so that the levels inside the quantum dot are not significantly broadened by tunneling. (b) A zoom-in scan of the boxed region in (a), revealing a series of resonant tunneling peaks. The resonant condition is found when source-drain Fermi level aligns with the quantized levels in the gate-defined quantum dot as shown in the schematic diagram at the top. The symbols correspond to the two different regimes with small (hexagon) and large (uptriangle) tunneling couplings. The separation and the slope of transitions (~1) stay reasonably constant over more than 10 consecutive charge transitions, implying a uniform dot being established near the center of the device with well-defined shape and charging energy. By tuning the tunnel couplings with the gate voltages, the quantized levels in the dot can be sensitively tuned from the thermally- broadened regime to the tunnel-broadened regime. (c) A charging energy of ~ 2 meV and a gate capacitive coupling ratio of ~ 0.15 can be extracted from the Coulomb diamonds (dashed lines), consistent with electrostatic simulations. Data are extrapolated in order to remove occasional background noise.

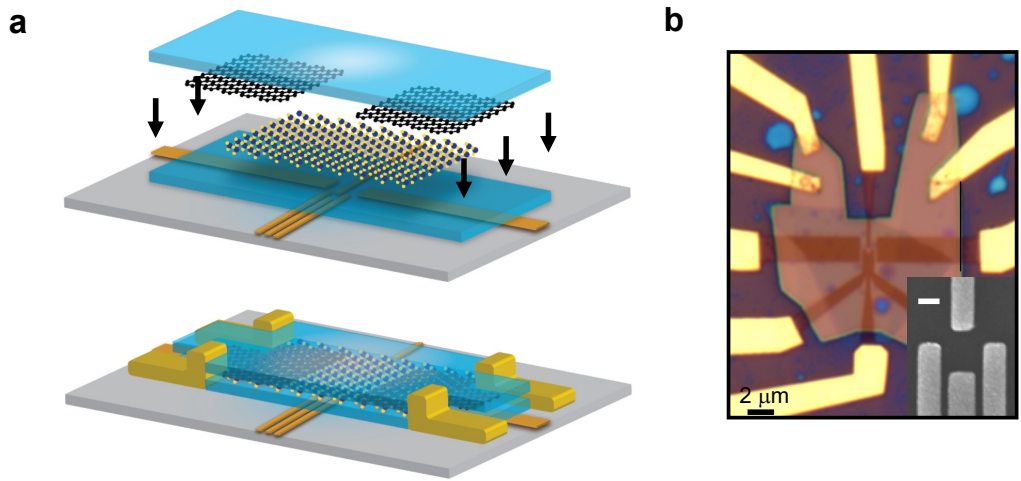

**Figure 1. hBN-graphene-MoS₂ heterostructure with local backgates**. (a) The van der Waals heterostructure consists of two hBN (~60nm-80nm), two single layer graphene and one trilayer MoS$_2$. The stack has been assembled and transferred on top of the pre-patterned gates, using the dry-transfer technique. Subsequently, Cr/Pd/Au edge contacts are made to graphene flakes, making van der Waals electrical contacts to the MoS$_2$ channel. (b) Optical microscope image of the gate-defined quantum dot device. The inset shows an scanning electron microscope image. The scale bar is 200 nm. Graphene contacts to MoS$_2$ are improved by positively-biased local contact gates, while the quantum confinement in the quantum dot is defined by negatively-biased local fine gates.

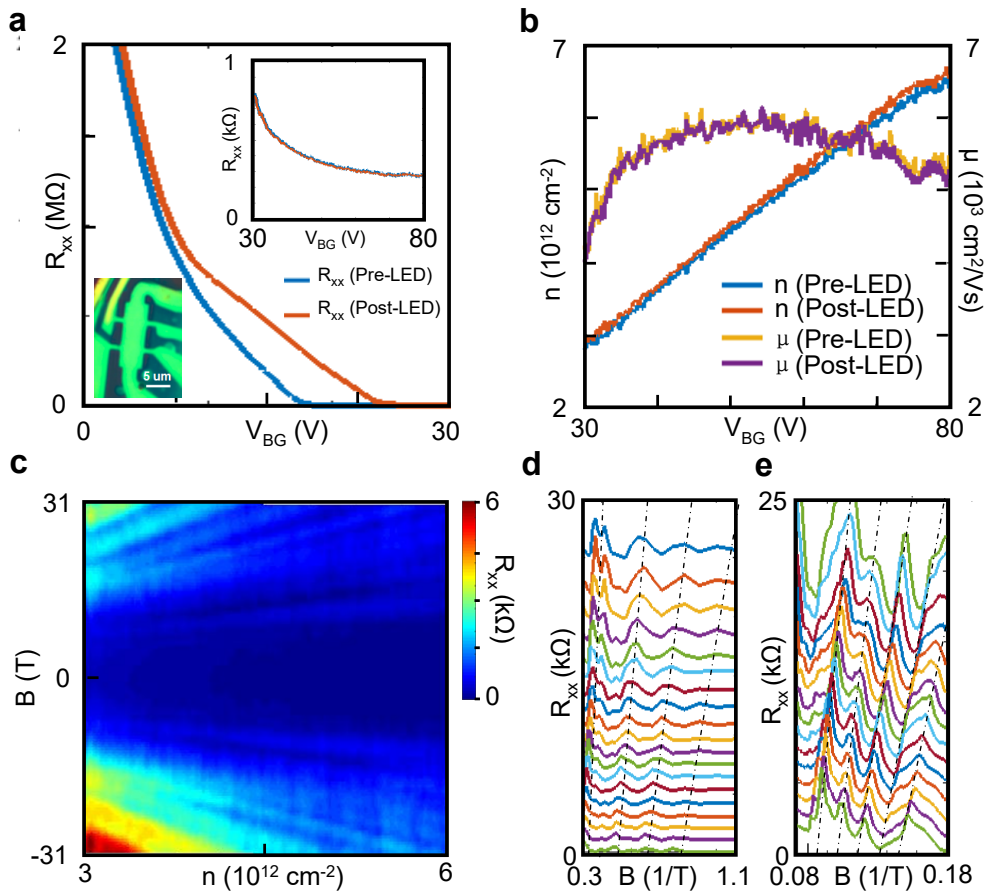

**Figure 2. Homogeneous TMDC 2DEG with low carrier density.** (a) Metal-insulator transition of a Hall bar made on trilayer $MoS_2$ encapsulated in Ar environment. The turn-on threshold voltage shifts to more positive values after flashing the device with red LED at 1.7 K. (inset) Sample resistance at higher backgate voltages show no noticeable difference before and after LED flash. (b) A large range of gate-tunable carrier density and a Hall mobility up to ~ 6,000 $cm^2$/Vs has been achieved, independent of the use of LED flash. (c) With transparent contacts achieved at a relatively low bulk carrier density, measurement of Landau level fan diagram is possible. (d) Low-field SdH oscillation of device made with vdW ion-implantation method (see SI). The traces are displaced vertically for clarity of presentation. The oscillation starting at ~1T implies the SdH mobility of ~ 10,000 $cm^2$/Vs, consistent with Hall measurement. The SdH peak near B=3T develops into two peaks, corresponding to partial symmetry breaking. (e) High magnetic field data shows data with periodicity that is four times faster in 1/B, possibly implying full symmetry-breaking of Landau levels or parallel conducting channels from each atomic layers.

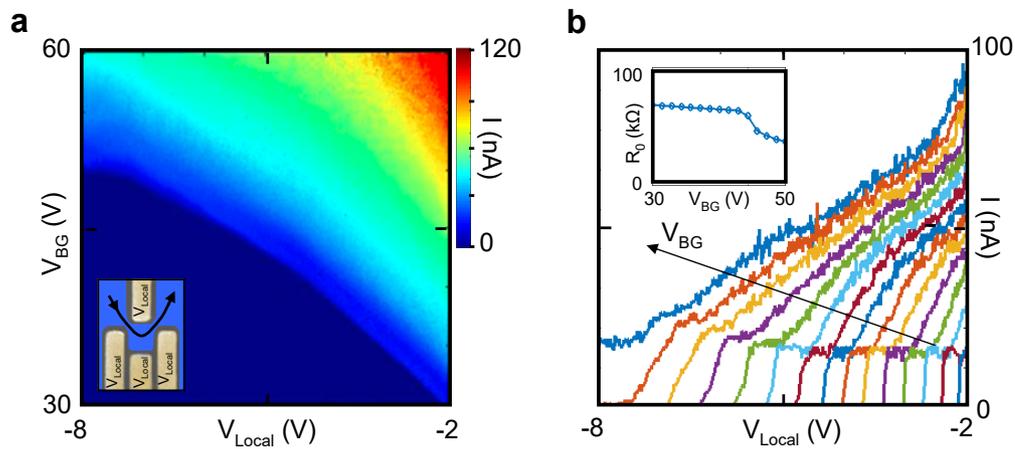

**Figure 3. Quantized conductance in quantum point contact.** (a) By energizing all the quantum dot gates ($V_{Local}$), two quantum point contact channels can be pinched off over a large range of electron density with a high on-off ratio exceeding $10^6$. The conductance changes monotonically, implying an uniform QPC channel with straight-forward gate-control. (b) 1D cuts of the pinch-off curve at various $V_{BG}$ (from 30 V to 52.5 V, with an equidistance step of 1.5 V) reveal development of quantum plateau, a signature of quantum confinement. The inset shows the series resistance to the QPC estimated from the quantized plateaus.

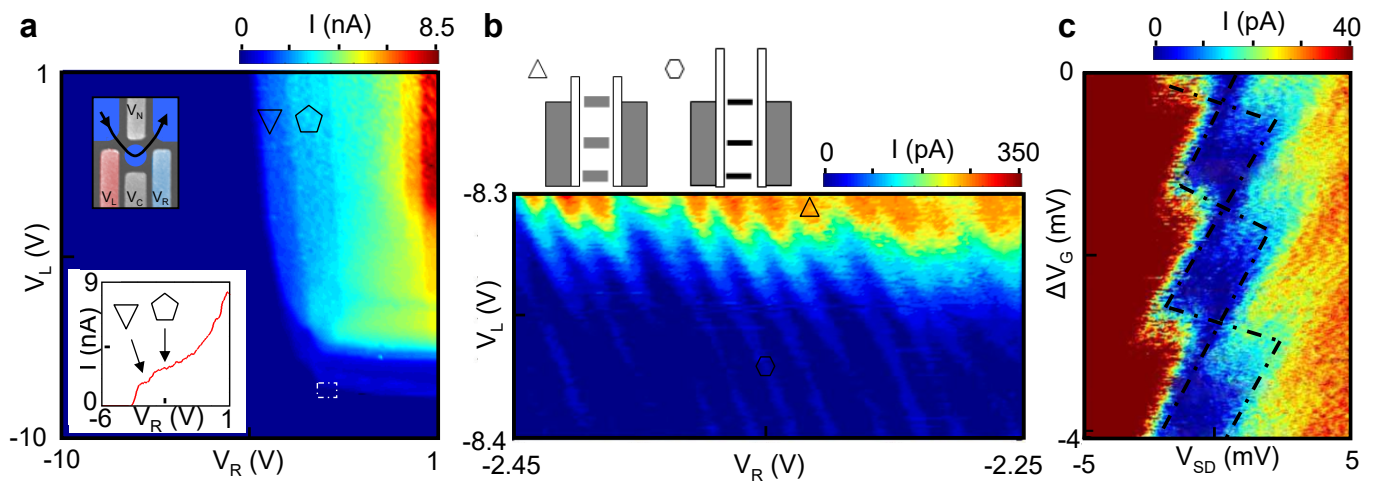

**Figure 4. Gate-defined quantum dot in MoS$_2$.** (a) Zero-bias ac conductance across the device as a function of left gate voltage $V_L$ and right gate voltage $V_R$ at $V_N$ = -8V and $V_C$ =-10V. The rectangular shape of the conductance region indicates that the left (right) tunnel coupling getting tuned by $V_L$ ($V_R$) independently, with visible quantum plateaus (inset). At the corner of the conducting region (dashed box), the tunnel couplings to the both contacts are optimized, so that the levels inside the quantum dot are not significantly broadened by tunneling. (b) A zoom-in scan of the boxed region in (a), revealing a series of resonant tunneling peaks. The resonant condition are is found when source-drain Fermi level aligns with the quantized levels in the gate-defined quantum dot as shown in the schematic diagram at the top. The symbols correspond to the two different regimes with small (hexagon) and large (uptriangle) tunneling couplings. The separation and the slope of transitions (~1) stay reasonably constant over more than 10 consecutive charge transitions, implying a uniform dot being established near the center of the device with well-defined shape and charging energy. By tuning the tunnel couplings with the gate voltages, the quantized levels in the dot can be sensitively tuned from the thermally-broadened regime to the tunnel-broadened regime. (c) A charging energy of ~2meV and a gate capacitive coupling ratio of ~ 0.15 can be extracted from the Coulomb diamonds (dashed lines), consistent with electrostatic simulations. Data are extrapolated in order to remove occasional background noise.